\documentclass[11pt]{article}

\makeatletter
\def\section{\@startsection {section}{1}{\z@}{-3.5ex plus -1ex minus
 -.2ex}{2.3ex plus .2ex}{\large\bf}}
\def\subsection{\@startsection{subsection}{2}{\z@}{-3.25ex plus -1ex 
minus -.2ex}{1.5ex plus .2ex}{\normalsize\bf}}
\makeatother
\makeatletter

\@addtoreset{equation}{section}

\makeatother

\font\cmss=cmss10 \font\cmsss=cmss10 at 7pt

\def\inbar{\vrule height1.5ex width.4pt depth0pt}
\def\IC{\relax\,\hbox{$\inbar\kern-.3em{\rm C}$}}
\def\IG{\relax\,\hbox{$\inbar\kern-.3em{\rm G}$}}
\def\IB{\relax{\rm I\kern-.18em B}}
\def\ID{\relax{\rm I\kern-.18em D}}
\def\IL{\relax{\rm I\kern-.18em L}}
\def\IF{\relax{\rm I\kern-.18em F}}
\def\IH{\relax{\rm I\kern-.18em H}}
\def\II{\relax{\rm I\kern-.17em I}}
\def\IN{\relax{\rm I\kern-.18em N}}
\def\IP{\relax{\rm I\kern-.18em P}}
\def\IQ{\relax\,\hbox{$\inbar\kern-.3em{\rm Q}$}}
\def\bfzero{\relax\,\hbox{$\inbar\kern-.3em{\rm 0}$}}
\def\IK{\relax{\rm I\kern-.18em K}}
\def\IG{\relax\,\hbox{$\inbar\kern-.3em{\rm G}$}}
 \font\cmss=cmss10 \font\cmsss=cmss10 at 7pt
\def\IR{\relax{\rm I\kern-.18em R}}
\def\ZZ{\relax\ifmmode\mathchoice
{\hbox{\cmss Z\kern-.4em Z}}{\hbox{\cmss Z\kern-.4em Z}}
{\lower.9pt\hbox{\cmsss Z\kern-.4em Z}}
{\lower1.2pt\hbox{\cmsss Z\kern-.4em Z}}\else{\cmss Z\kern-.4em
Z}\fi}
\def\bfone{\relax{\rm 1\kern-.35em 1}}

\def\bar{\overline}

\def\IE{\relax{{\rm I\kern-.18em E}}}

\def\IGam{\relax{{\rm I}\kern-.18em \Gamma}}

\def\beq{\begin{equation}}
\def\eeq{\end{equation}}
\def\beqa{\begin{eqnarray}}
\def\eeqa{\end{eqnarray}}
\def\nn{\nonumber}

\begin{document}

\begin{titlepage}
\setcounter{page}{0}

\begin{flushright}

NORDITA--2000/03 HE

SWAT/256

\end{flushright}

\vskip 20pt

\begin{center}

{\Large \bf Regular BPS black holes: macroscopic and microscopic \\  
description of the generating solution}

\vskip 20pt

{\large Matteo Bertolini}\footnote{Supported by INFN, Italy.} 
\vskip 5pt
{\it NORDITA \\ Blegdamsvej 17, DK-2100 Copenhagen \O, Denmark} \\
{\footnotesize{teobert@nordita.dk, teobert@nbi.dk}}
\vskip 7pt
and
\vskip 7pt
{\large Mario Trigiante}
\vskip 5pt
{\it Department of Physics, University of Wales Swansea \\ 
Singleton Park, Swansea SA2 8PP, United Kingdom} \\
{\footnotesize{m.trigiante@swan.ac.uk}}

\end{center}

\vskip 20pt

\begin{abstract}

In this paper we construct the BPS black hole generating solution of toroidally 
compactified string (and $M$) theory, giving for it {\it both} the macroscopic 
and microscopic description. Choosing a proper $U$--duality gauge the latter will 
be given by a bound state made solely of D--branes. The axionic nature of the 
supergravity solution will be directly related to non--trivial angles between 
the constituent D--branes (type IIB configuration) or, in a $T$--dual gauge, 
to the presence of magnetic flux on constituent D--brane world volumes (type IIA 
configuration). As expected, the four dimensional axion fields arise from the 
dimensional reduction of non--diagonal metric tensor components or Kalb-Ramond B field 
components for type IIB or type IIA cases, respectively. Thanks to this result it is 
then now possible to fill the {\it full} 56--dimensional $U$--duality orbit of $N=8$ 
BPS black holes and to have a macroscopic and microscopic description of all of them.
\end{abstract}

\end{titlepage}

\tableofcontents


\section{Introduction and summary of the results}

In the last 4 years the study of black hole solutions in supergravity
and string theory has acquired a renewed interest. This is due,
essentially, to some successful microscopic entropy  countings which
have given, for the first time \cite{strvaf,malstr,jkm,lowe}, a 
statistical interpretation of the Beckestein--Hawking entropy formula
in the context of a consistent quantum theory of gravity, such as string
theory. The general properties of both supersymmetric and near
supersymmetric black holes  have been systematically studied and
many new results have been obtained. 

Despite this  progress, it is still not possible  nowadays to have a 
complete control on the common microscopic properties underlying  these 
black holes. Many efforts have been made in order to find a 
way of making these microscopic entropy countings more based on first 
principles rather than on case by case computations (see for example 
\cite{msw,str1,porn,mand,carlip,gt,moooa,finn1}) and to understand 
which are the actual microscopic degrees of freedom underlying the 
microscopic/macroscopic matching. However, a definite answer has not been found, yet.
In this respect, it would be useful to find a precise description, {\it both} 
at macroscopic and microscopic level, of the so--called generating solution 
of regular BPS black holes obtained in the framework of superstring (or $M$) 
theory compactified on the torus $T^6$ ($T^7$).

There are at least three known macroscopic generating solutions in the 
literature, \cite{cve1,cve2,bft2}. However, their structure  is too involved 
to allow a direct construction of the corresponding microscopic
configuration. On the other hand, it was proposed in \cite{bala2}
one possible stringy (i.e. microscopic) structure for the generating
solution but it was not possible to find its field theory description
as a solution of the relevant supergravity theory.

In this paper we fill the above gap and finally find the $N=8$ BPS black
holes  generating solution, giving for it {\it both} a macroscopic
and a microscopic description in a clear and simple way. As expected,
the solution will depend on five independent parameters which will
have a precise physical meaning both at macroscopic and microscopic
level. Despite the number of independent parameters is five, the
number of independent harmonic functions entering the solution  will
be four. This is somewhat expected and is related to the well known fact
that within the five invariants of the $U$--duality group of 
toroidally compactified $M$--theory, i.e. $E_{7(7)}$, four are moduli--dependent  
and one is moduli--independent (and proportional to the entropy). Another 
 feature of the generating solution is to have non--trivial axion fields switched 
on in it (this is related, as we shall see, to the intrinsic property of the 
central charge eigenvalues evaluated on the solution of being intrinsically complex, i.e. 
not all imaginary). On the microscopic side this will correspond to have a 
configuration of D--branes intersecting at non--trivial angles (IIB) or, in a 
$T$--dual picture, a D--brane configuration with non--trivial magnetic fluxes on the 
D--brane world--volumes (IIA). As expected, the non--trivial four dimensional 
axions will come from off--diagonal components of the ten dimensional metric $G_{MN}$ 
in the type IIB case and of the antisymmetric tensor $B_{MN}$ in the type IIA case, along the world 
volume of the D--branes (and constant with respect to the world volume coordinates). 
Switching off the magnetic flux one would recover a four parameter solution of D--branes 
orthogonally intersecting, which turns out to be a pure dilatonic one.

Once a macroscopic solution it is given, its  specific
microscopic counterpart it is not uniquely defined: it depends on how
the solution is embedded in the original ten dimensional theory. Our choice is 
to use two suitable R--R embeddings recently defined in \cite{mm}, in order to have, 
as anticipated, a stringy configuration made solely of D--branes. 
Relying on the structure of the  superalgebra central charges and thanks to 
the geometrical control of the embedding of the solution in the full type II 
string theory, it will be quite  easy to figure out the microscopic systems  
corresponding to our solution. Moreover, using the tools developed in previous papers 
(\cite{bft2,mp1,bft} and especially \cite{mm}) we are able to generate, from 
the above simple pure D--brane description, any other configuration, even pure 
NS--NS ones, i.e. those made of solely NS states (as fundamental strings, 
NS5--branes and KK--states)\footnote{As shown in \cite{cvhu}, the generating 
solution of heterotic black holes is also a generating solution for the type II 
ones. It is the $U$--duality group which changes in the two cases and 
which specifies the $U$--duality properties of the solution. In the heterotic 
case is $U=SL(2)\times SO(6,22)$ while in the present case, i.e. type IIA, 
type IIB or $M$--theory compactified on tori ($T^6$ or $T^7$, respectively), 
we have $U=E_{7(7)}$.}. 
The property for the entropy of being an $U$--duality invariant ensures that all 
these $U$--dual configurations share the same entropy. And, as already stressed, 
the possibility of having a control both at macroscopic and microscopic level of 
all these configurations, could give some help in unraveling the very conceptual 
basis of the microscopic entropy counting.

\section{The macroscopic generating solution}

Let us start with the macroscopic description of the generating
solution. It has been shown in \cite{mp1} that the BPS black
holes generating solution of toroidally compactified type  II string
is also solution of a  consistent truncation of the relevant $N=8$
supergravity  effective theory, the so called $STU$ model. 
This is a $N=2$ effective  model characterized by the graviton multiplet 
and just three vector multiplets, each of them containing two scalar
fields\footnote{ BPS solutions of $N=2$ supergravity have been studied, for instance, 
in \cite{shma,behr,cve3,ferka,kal1,sabra1}.}. Therefore its full bosonic field content 
is: a graviton, four vector fields (i.e. eight charges, ($p_\Lambda,q_\Lambda$) where 
$\Lambda=0,1,2,3$) and three complex scalars $z^i=a_i+{\rm i} b_i$ spanning the 
Special K\"ahler manifold ${\cal M}_{STU}=\left[SL(2,\IR)/SO(2)\right]^3$
($b_i$ being the dilatonic fields and $a_i$ the axions).
 We are not going to describe
the detailed structure of the model since it has been described in a
complete way in \cite{bft2,bft}. In fact, we will use the same
conventions and notations adopted in those papers. 
Let us then just briefly summarize the procedure to follow in order to derive the
 generating solution in the framework of the $STU$ model, leaving to the next 
section the discussion of its embedding in the $N=8$ theory.
The BPS condition is equivalent to imposing the vanishing of the fermion
supersymmetry variation along the {\it Killing spinor} $\xi_a$  direction:
\begin{eqnarray}
\delta_\xi\mbox{fermions}&=&0 \nn \\  \gamma^0 \,\xi_{a} &=&  \pm
\frac{Z}{|Z|}\,\epsilon_{ab} \, \xi^{b} \quad \mbox{if} \quad a,b=1,2
\label{killing}
\end{eqnarray}
$ Z(z,\bar{z},p,q) $ being the $N=2$ supersymmetry central charge.
We adopt the following ans\"atze for the metric, Killing spinor, scalars and vector 
field--strengths:
\begin{eqnarray}
ds^2&=&e^{2{\cal U}\left(r\right)}dt^2-e^{-2{\cal
U}\left(r\right)}d\vec{x}^2 ~~~~~
\left(r^2=\vec{x}^2\right)\nonumber\\ 
\xi_a(x)&=& f(r)\epsilon_a\nonumber\\
 z^i(x)&=&z^i(r) \nn \\
F^\Lambda(r)&=&\frac{p_\Lambda}{2 r^3}\epsilon_{krs} x^k dx^r \wedge
x^s - \frac{l_\Lambda(r)}{r^3} e^{2{\cal U}(r)} dt \wedge \vec{x} \cdot
d\vec{x} \quad,\quad \Lambda=0,1,2,3
\label{fishlicense}
\end{eqnarray}
$l_\Lambda$ being the moduli--dependent electric charges defined in \cite{bft}.
From the BPS conditions (\ref{killing}) we may derive an equivalent system of first 
order equations for the scalars and metric function ${\cal U}$:
\begin{eqnarray}
\frac{dz^i}{dr}\, &=&\, \mp 2\left(\frac{e^{{\cal U}(r)}}{r^2}\right)
h^{ij^\star}\partial_{j^\star} \vert Z(z,\bar{z},{p},{q})\vert \nn\\
\frac{d{\cal U}}{dr}\, &=&\, \mp \left(\frac{e^{{\cal
U}(r)}}{r^2}\right)  \vert Z(z,\bar{z},{p},{q})\vert
\label{eqs122}
\end{eqnarray}
The explicit expression of the right hand side of eq.s (\ref{eqs122})
is quite involved and may be derived from the equations in \cite{bft}
(computed using the too restrictive condition  $Z=-\bar{Z}$) by
multiplying their right hand side by $\sqrt{Z/\bar{Z}}$ or by
$\sqrt{\bar{Z}/Z}$ as far as the equations for the scalars or the one
for ${\cal U}$ are concerned, respectively. 

In order to characterize
the generating solution we need to compute the skew--eigenvalues
$\{Z_\alpha\}=\{Z_{\hat{i}},Z_4\}$ ($\alpha=1,\dots,4$) of the $N=8$ central charge $Z_{AB}$
in terms of the supersymmetry and {\it matter} central charges $\{Z,
Z_i\}$ of the $N=2$ model, using the following relations: 
\beqa
\label{stun8}
Z\,=\,i\,Z_4 \quad,\quad
Z^i\,=\,h^{ij^\star}\nabla_{j^\star}\bar Z\,=\,\IP^i_{\hat{i}} Z^{\hat{i}}\,\,\,\,\,\,\,\,
(i,{\hat{i}}=1,2,3)
\eeqa 
where $\IP^i_{\hat{i}}= 2 b_i(r)$ is the vielbein transforming the
rigid indices ${\hat{i}}$ (the one  characterizing the eigenvalues of the
$N=8$ central charge in its normal form) to the curved indices  $i$ of the
$STU$ scalar manifold (see \cite{bft}, section 3, for details). Following 
\cite{bft}, the explicit expression of the $N=8$ central charge eigenvalues in 
terms of quantized charges and moduli is given in the appendix. The five
$U$--duality invariants characterizing a generic solution are the four
norms $\vert Z_\alpha \vert $ and the overall phase $\Phi=\sum_\alpha
{\rm Arg}(Z_\alpha)$, \cite{fermal}. These invariants may be combined to give four
moduli--dependent invariants and one moduli--independent invariant (the
quartic invariant of the $U$--duality group). The generating solution
is defined \cite{bft2} as the BPS black hole solution depending on the
least number of parameters such that, on the point of the moduli space
$\phi^\infty$ defining the boundary condition at radial infinity of
its scalar fields, the five invariants can assume all $5$--plets of
values (consistent with the positivity condition of the quartic
invariant). A necessary condition for the generating solution is thus
to depend only on five quantized charges, obtained from the original 8
by suitably fixing the $SO(2)^3$ gauge.  Our choice for the gauge
fixing is $p_0=q_2=q_3=0$. As an a--posteriori check that the solution
is a generating one, it is necessary to verify that the five invariants
computed in the corresponding $\phi^\infty$ are independent functions of the
five remaining charges: $q_0,q_1,p_1,p_2,p_3$. 

With the above gauge choice the system of first and second order differential 
equations symplifies considerably and the fixed values for the scalar fields
(namely the values the scalars get at the horizon, \cite{fer}) turn
out to be the following ones:    \beqa
&&a_1^{fix}\,=\,-\frac{q_1p_1}{2 p_2p_3}\quad,\quad b^{fix}_1\,=\,
-\sqrt{\frac{q_0p_1}{p_2p_3}-\frac14\left(\frac{q_1p_1}{p_2p_3}\right)^2}\nn\\
&&a_2^{fix}\,=\,\frac{q_1}{2 p_3}\quad,\quad b^{fix}_2\,=\,
-\sqrt{\frac{q_0p_2}{p_1p_3}-\frac 14 \left(\frac{q_1}{p_3}
\right)^2}\nn \\   &&a_3^{fix}\,=\,\frac{q_1}{2 p_2}\quad,\quad
b^{fix}_3\,=\, -\sqrt{\frac{q_0p_3}{p_1p_2}-\frac 14
\left(\frac{q_1}{p_2} \right)^2} \eeqa   Let us now introduce the
following harmonic functions:
\begin{eqnarray}
&&H^i(r)\,=\,1+\frac{\sqrt{2}p_i}{r} \quad \mbox{with}\quad i=1,2,3 \nn \\
&&H_0(r)\,=\,1+\frac{\sqrt{2}q_0}{r} \quad \mbox{and}\quad 
H_1(r)\,=\,g+\frac{\sqrt{2}q_1}{r}\;\;,\;\; g\equiv\frac{q_1}{p_1+p_2}
\label{armf}
\end{eqnarray}
where the above value for the parameter $g$ is fixed by supersymmetry 
(first order equation for ${\cal U}$).\par
One can now see that the following ans\"atze for the $a_i(r)$, the $b_i(r)$ and
the scalar function ${\cal U}(r)$:
\begin{eqnarray}
&&a_1\,=\,\frac{-H_1H^1+gH^2}{2 H^2H^3}\;,\;\; b_1\,=\,
-\sqrt{\frac{H_0H^1}{H^2H^3}-\frac14\left(\frac{H_1H^1-gH^2}{H^2H^3}\right)^2}\nn
\\ &&a_2\,=\,\frac{H_1H^1-gH^2}{2 H^1H^3}\;,\;\; b_2\,=\,
-\sqrt{\frac{H_0H^2}{H^1H^3}-\frac14\left(\frac{H_1H^1-gH^2}{H^1H^3}\right)^2}\nn
\\ &&a_3\,=\,\frac{H_1H^1+gH^2}{2 H^1H^2}\;,\;\; b_3\,=\,
-\sqrt{\frac{H_0H^3}{H^1H^2}-\frac14\left(\frac{H_1H^1-gH^2}{H^1H^2}\right)^2}\nn
\\ &&{\cal U}\,=\,-\frac{1}{4}\ln\left(H_0H^1H^2H^3
-\frac{1}{4}(H_1H^1-gH^2)^2\right)
\label{ans}
\end{eqnarray}
satisfies both the first and second order differential equations and
hence is the solution we were looking for\footnote{This result is consistent with 
the analysis in \cite{sabra1}.}. The values of the constants 
characterizing each harmonic function have been chosen in such a way to have 
1) asymptotic flat space and 2) asymptotic unitary values for the dilatons $b_i$, 
so to have unitary radii of compactification. The corresponding point in the moduli 
space at infinity is thus:
\begin{eqnarray}
\phi^\infty &\equiv &\cases{a_1=a_2=0;\,a_3=g\cr b_i=-1}
\label{phinfty}
\end{eqnarray}
Since the boundary values of the scalar fields at infinity define a bosonic 
vacuum of the theory, they characterize also the microscopic configuration realizing 
our solution in the opposite string coupling regime. In next section we shall describe 
two particularly simple microscopic configurations corresponding to
the choice of $\phi^\infty$ in eq. (\ref{phinfty}).

Notice that the  number of truly independent harmonic functions in eq.(\ref{ans}) is four, 
as expected, although the number of independent charges is five: $q_0,q_1,p_1,p_2,p_3$. 
Indeed, from the 
conditions (\ref{armf}), one sees that $H_1(r)\,=\,g\,\left(H^1(r)+H^2(r)-1\right)$. Finally, 
according to the ans\"atze (\ref{fishlicense}), the metric has the following form:  
\beqa  
\hskip -8pt
ds^2&=&\left(H_0H^1H^2H^3 -\frac{1}{4}(H_1H^1-gH^2)^2\right)^{-1/2}dt^2- \left( H_0H^1H^2H^3
-\frac{1}{4}(H_1H^1-gH^2)^2\right)^{1/2} d\vec{x}^2 \nn \\   
\eeqa 
and the macroscopic entropy, according to Beckenstein--Hawking formula,
reads:   
\beq
\label{entrqp}
S_{macro}= 2\,\pi\,\sqrt{q_0p_1p_2p_3 -\frac{1}{4}(q_1p_1)^2}   
\eeq
which is the expected expression for the entropy of a generating solution, \cite{cve1,cve2}. 

As anticipated, in order to check that the above solution is indeed a five 
parameters one, one has to work out the expression of the $N=8$ central charge 
skew--eigenvalues on $\phi^\infty$. From the explicit expressions of the field dependent 
central charge in the appendix and from (\ref{phinfty}) it follows that:
\beqa  
Z_1(\phi^\infty,p,q)&=&\frac{1}{2\sqrt{2}}\left[-2\frac{q_1p_1}{p_1+p_2}  +
i\left(q_0 + p_1 - p_2 - p_3 \right)\right]\nn\\
Z_2(\phi^\infty,p,q)&=&\frac{1}{2\sqrt{2}}\left[2\frac{q_1p_1}{p_1+p_2} + i\left(q_0 -
p_1 + p_2 - p_3\right)\right]\nn\\ 
Z_3(\phi^\infty,p,q)&=&\frac{1}{2\sqrt{2}}\left[0 +
i\left(q_0 - p_1 - p_2 + p_3 \right)\right]\nn\\
Z_4(\phi^\infty,p,q)&=&\frac{1}{2\sqrt{2}}\left[0  + i\left(q_0 + p_1 + p_2 +
p_3\right)\right]
\label{chargeqp}
\eeqa
From the above equations it is clear that the 5 invariant quantities 
$\vert Z_\alpha (\phi^\infty,p,q)\vert$, $\Phi(\phi^\infty,p,q)$ 
are independent functions of the five charges $q_0,q_1,p_1,p_2,p_3$.

Switching off the fifth parameter $q_1$ our solution becomes
exactly the four parameters one studied in \cite{mm} (indeed
$H_1(r)\vert_{q_1=0}=0\;\mbox{and}\;g(q_1=0)=0$). Indeed putting $q_1$
to zero the central charges $Z_\alpha $ become pure imaginary and 
the axion fields vanish uniformly.

\section{The microscopic description of the generating solution} 

The microscopic counterpart of a four dimensional macroscopic solution
 is of course not uniquely defined. Indeed it depends on the
 interpretation of  the four dimensional fields describing the
 supergravity  solution in terms of dimensionally reduced ten
 dimensional ones. In the light of the analysis put forward in
 \cite{solv} and then completed in \cite{mm}\footnote{Among the main goals 
of this analysis is the geometrical characterization, using solvable Lie algebra (SLA)
 techniques, of the scalar and vector fields in the $N=8$ theory in terms
 of type IIA and IIB fields.}, the
 microscopic interpretation of our generating solution can be uniquely
 defined in terms of the embedding of the $STU$ model within the original
 $N=8$ theory (in particular of the embedding of the SLA generating the
 $STU$ scalar manifold inside the SLA parametrized by the $70$ scalars
 of the $N=8$ theory). In \cite{mm}, two main classes of embeddings
 of the $STU$ model were defined (the embeddings within each class being 
related by $S,T$ dualities):  one
 in which the vector fields derive from NS--NS ten dimensional forms
 and the other in which the vector fields have a R--R origin ( and the
 scalar fields  are NS--NS, see in particular section 2 of that
 paper).  In the latter class two representative embeddings were
 considered: if we  denote by $x^4\,,\,x^5\,,...,\,x^9$ the
 coordinates of $T^6$ and by $x^0\,,\,x^1\,,\,x^2\,,\,x^3$ the
 non--compact space--time coordinates, in one  embedding, which was
 characterized from the type IIB point of view, all the three  axions of
 the model come from metric tensor components ($G_{45},G_{67},G_{89}$), 
as opposite to the other, characterized from a type IIA point of view, in which 
all the axions come from $B$ field components ($B_{45},B_{67},B_{89}$). In both 
cases the three dilatons are related to three combinations of the radii of the torus. The two 
embeddings are related by an operation of $T$--duality on the compact 
directions $x^5,x^7,x^9$. As we shall show in the sequel these two embeddings
 provide an  interpretation of the generating solution (\ref{ans}) in
 terms of two $T$--dual  microscopic  configurations: a system of
 D3--branes at angles (type IIB embedding) and  a system of D0 and
 D4--branes (type IIA embedding) with a magnetic flux in the
 world  volume of the latter (giving therefore extra D2 and D0 charge,
 \cite{doug}). The magnetic flux (or, equivalently, the non--trivial angle 
in the dual type IIB configuration) will be the microscopic extra degree of freedom 
related to the fifth parameter $q_1$ characterizing the supergravity  solution. 

Analyzing the two embeddings from a SLA point of view, one can deduce,
in the  same way as it was done in \cite{mm} for the type IIA case
(see in particular eq.s (3.14) and (3.15) of that paper), the subset of the 
weight basis of the ${\bf 56}$ of $E_{7(7)}$ in terms of which the {\it magnetic}  $y^n(\phi)$ 
($n=0,\dots,3$) and {\it electric} $x_n(\phi)$ {\it dressed  charges}\footnote{
The dressed charges are the physical charges of the interacting theory, which take 
into account the dressing of the D-brane naked charges provided by the moduli.} of the 
two $STU$ model 
truncations are expressed. According to \cite{mm}, the symplectic vector of 
dressed charges $(y^n,x_n)$ is defined in the following way:
\begin{eqnarray}
\left(\matrix{y^n(\phi)\cr x_n(\phi)}\right)&=&
-\IL^{-1}(\phi)\left(\matrix{p^n\cr q_n}\right)
\label{dressed}
\end{eqnarray}
where $\phi$ denotes a point in the scalar manifold and  $\IL(\phi)$
is the coset representative of the scalar manifold computed in the same 
point. As far as the type IIA embedding is concerned, a basis of weights for
$(y^n,x_n)$ was found in \cite{mm}, and from table 3 of the same work
the correspondent  R--R vectors may be read off:
\begin{eqnarray}
\mbox{type IIA}:&&\nonumber\\ (y^n)&\leftrightarrow& (A_{\mu
456789},A_{\mu 6789},A_{\mu 4589},A_{\mu 4567})\nonumber\\ 
(x_n) &\leftrightarrow& (A_{\mu},A_{\mu 45},A_{\mu 67},A_{\mu 89})
\label{iia}
\end{eqnarray}    
By performing a $T$--duality along $x^5,x^7,x^9$ according to the
geometric recipe given in \cite{mm}, we may find the corresponding weights 
for the type IIB embedding and
read from table 3 of  the same work their R--R interpretation:
\begin{eqnarray}
\mbox{type IIB}:&&\nonumber\\ (y^n)&\leftrightarrow& (A_{\mu
468},A_{\mu 568},A_{\mu 478},A_{\mu 469})\nonumber\\ 
(x_n) &\leftrightarrow& (A_{\mu 579},A_{\mu 479},A_{\mu 569},A_{\mu 578})
\label{iib}
\end{eqnarray}
Now let us consider our generating solution and compute the dressed
charges on the point of  the moduli space $\phi^\infty$ defined in eq. (\ref{phinfty}). 
Implementing eq.(\ref{dressed}), one finds: 
\begin{eqnarray}
(y^0,y^1,y^2,y^3)&=&(0,-p_1,-p_2,-p_3)\nonumber\\
(x_0,x_1,x_2,x_3)&=&(-q_0,-\frac{p_1q_1}{p_1+p_2},\frac{p_1q_1}{p_1+p_2},0)
\label{ourdress}
\end{eqnarray}
From the above expressions we may deduce consistent microscopic
configurations corresponding to the generating solution with the
chosen boundary condition on the scalar fields at infinity
$\phi^\infty$. From the type IIB viewpoint we may think of a system of 
D3--branes intersecting at non--trivial angles, but in such a way to preserve 
1/8 supersymmetry; this can be achieved if the relative rotation between each couple 
is a $SU(3)$ rotation, \cite{jab}. The configuration is depicted in table \ref{ND3brane}.

\begin{table} [ht] 
\vskip 10pt
\begin{center}
\begin{tabular}{|c|c|c|c|}
\hline & $\phi_1$ & $\phi_2$ & $\phi_3$ \\  
\hline $N_0$ & $\pi/2$  & $\pi/2$  & $\pi/2$ \\  
\hline $N_1$ & $\pi/2$  & 0  & $\pi$  \\  
\hline $N_2$ & $\pi$  & $\pi/2$  & 0 \\  
\hline $N_3$ & $\theta$ & $\pi-\theta$ & $\pi/2$ \\  
\hline
\end{tabular}
\end{center}
\caption{\small The position of the D3--branes on the compactifying
torus; $\phi_i$ is the angle on the $(x^{2i+2},x^{2i+3})$ torus and 
$\theta$ is a generic non--trivial angle. For each couple of constituent D3-branes, 
it follows that $\sum_{i=1}^3 (\phi^{(\alpha)}_i-\phi^{(\beta)}_i)=0\;
\mbox{mod}\;2\pi\;\,\forall \alpha,\beta=1,...,4$, this ensuring that it is a 
configuration of 4 (bunches of) D3--branes at $SU(3)$ angle, \cite{jab,bala0}. 
Notice that the above configuration has been chosen in such a way that setting 
$\theta=0$ one recovers a four parameters solution, namely $4$ bunches of 
D3--branes orthogonally intersecting.}
\label{ND3brane}
\end{table}

Using eq.s (\ref{iib}) the first three sets of branes ($N_0,N_1,N_2$) may be associated 
with the charges $x_0,y^1,y^2$ respectively, i.e. with charge along the 3--cycles 
$(579)$, $(568)$, $(478)$, while the fourth set, $N_3$, with $y^3$, i.e. the charge along 
$(469)$. In fact, due to the non--trivial angle $\theta$,  the fourth set of $N_3$ branes 
induces D3--brane charge on the cycles $(579)$ (contributing to $x_0$), $(479)$ 
(represented by $x_1$) and $(569)$ (represented by $x_2$).
 
As far as the type IIA microscopic interpretation is concerned, we may 
consider the configuration of D0 and D4--branes obtained by $T$--dualizing 
the type IIB one described above along the directions $x^5,x^7,x^9$. The 
corresponding system may be deduced from eq.s (\ref{iia}) and (\ref{ourdress}) 
and consists of a set of coinciding D0--branes with electric charge $-x_0$ (the minus sign 
is required by consistency with the construction in \cite{mm} and will be discussed in the sequel) 
and three sets of coinciding D4--branes along the four--cycles $(6789)$, 
$(4589)$ and $(4567)$ with magnetic charges $y^1,y^2,y^3$. In addition there 
is a magnetic flux (related to the angle $\theta$ in the T--dual type 
IIB configuration, \cite{li,ck}) switched on the world volume of the latter brane (i.e. 
along $(4567)$). This flux induces an effective D0 charge (contributing to $x_0$) 
and effective D2 charges along the two--cycles $(45)$ and $(67)$ (represented by $x_1$ 
and $x_2$, respectively). The presence of this flux is also consistent with the 
fact that the axions in the type IIA embedding are interpreted as coming from the $B_{MN}$ tensor 
in ten dimensions. Indeed, let us briefly recall the general argument relating the presence of 
a flux on one D4--brane with an effective D2--brane charge (electric in our framework) 
and a non--trivial $B_{ij}$ background field. As well known, $B$ field components enter 
non--trivially in the D$p$--brane action via the WZ term:
\beq 
\mu_p\int_{W_{p+1}}\left(C\wedge e^{{\cal F}}\right)_{p+1} 
\label{cerns}
\eeq 
where ${\cal F}$ is the gauge invariant combination ${\cal F}=2\pi\alpha' F + \hat B$ 
($\hat B$ being the pull--back of the $B$ field). 
Hence, from the supergravity point of view, one would indeed expect new charges 
representing extra D$(p-2)$ effective charges at the microscopic level as 
well as non--trivial bulk $B$ field components in the solution 
(see for instance \cite{gar}). In fact, this is precisely what we get. 
As shown for instance in \cite{bala0,bala1}, for a suitable choice of the flux (which, 
albeit giving smaller brane charges via world--volume Chern--Simons coupling, modifies 
the supersymmetry projections imposed by the D--brane background) the above configuration 
can preserve $1/8$ of the original supersymmetry. Consider the D4--brane configuration 
described above in the general situation in which the magnetic fluxes are non vanishing on 
all the three planes $(45)$, $(67)$, $(89)$ 
(i.e. ${\cal F}_{45},\,{\cal F}_{67},\,{\cal F}_{89}\neq 0$). From eq. (\ref{cerns}) we may 
deduce the effective (electric) D2 and D0--brane charges. For instance, the effective 
D2--brane charge along $(45)$ (which is the electric dual object of the D4--brane wrapped 
on $(6789)$) is:
\begin{eqnarray}
&&\mbox{\# of $D2$ brane (along cycle 45)}= \frac{1}{2\pi}
\left(\int_{T_{67}}{Tr\,{\cal F}_{67}}+ \int_{T_{89}}{Tr\,{\cal
F}_{89}}\right)
\label{numd2}
\end{eqnarray}
which in our conventions is represented by the electric dressed charge $x_1$.
Similarly, we may compute the effective $D2$ charges along the other two--cycles. 
Notice that the only non--vanishing components of ${\cal F}$ can only be 
${\cal F}_{45}\,,\,{\cal F}_{67}\,,\,{\cal F}_{89}$ (a ${\cal F}_{56}\neq 0$ 
component would imply, for instance, a new $4$ dimensional magnetic effective 
charge out of the four  at disposal in the  central charge normal gauge), and the three 
axions come precisely from those components of the $B$ field. 

On our particular solution the D2--brane charge along $(89)$, i.e. $x_3$, is zero, 
while $(45)$ and $(67)$ charges are opposite one to each other 
($x_1=-x_2$), see (\ref{ourdress}). In fact, considering eq.(\ref{numd2}) written also 
for the other 2--cycles, one can easily see that switching on a magnetic flux, as we do, 
only on the D4--branes lying along $(4567)$ and not on the other two bunches of D4--branes, 
is consistent with having no D2--brane charge along $(89)$, i.e. $x_3=0$, and 
opposite D2--brane charge along $(45)$ and $(67)$, i.e. $x_1=-x_2$. 
These charges turn out to be proportional to the {\it fifth parameter} $q_1$: 
sending $q_1$ to zero the fluxes vanish together with the $B$ fields 
(axions) and we recover the four parameter solution discussed in \cite{mm}. 

Let us now come to our final goal that is to make the macroscopic/microscopic correspondence 
precise, namely to give the precise matching between the parameters characterizing the 
microscopic and the macroscopic configurations, respectively:
\begin{eqnarray}
N_0\,,\,N_1\,,\,N_2\,,\,N_3\,,\,\theta \;\longleftrightarrow
\;q_0\,,\,q_1\,,\,p_1\,,\,p_2\,,\,p_3  \nn
\end{eqnarray}
The type IIA and IIB D--brane configurations discussed above as the 
microscopic counterparts of our generating solution, were suggested in \cite{bala2} 
as candidates for the microscopic representation of the 5--parameter 
solution, whose macroscopic description was then missing. Let us focus for the moment on the 
type IIA embedding. In order to make contact 
with this literature, let us use an equivalent representation of the dressed charges, 
related to the central charges by an $SO(8)$\footnote{The subgroup of $SU(8)$ which 
does not ``mix'' electric and magnetic charges.} transformation:
\beq 
z_{ij}=x_{ij}+iy^{ij}=-\frac{1}{\sqrt{2}}\,\left(\Gamma^{AB}\right)_{ij}\,Z_{AB}
\label{zZ}
\eeq  
where the couple $(ij)$ indicizes  the two times antisymmetric representation of $SO(8)$. 
The real and imaginary parts of $z_{ij}$ are the $N=8$ electric and magnetic dressed charges 
in the basis of weights of the ${\bf 56}$ of $E_{7(7)}$ defined in \cite{mm} and listed in 
table 3 of the same paper.

When $Z_{AB}$ is skew--diagonal the matrix $\Gamma\equiv \left(\Gamma^{AB}\right)_{ij}$ has 
the form:  
\beq  
\Gamma\, =\,\left( \begin{array}{cccc}  
\;\;\,1 & -1 & -1 & \;\;\,1 \\  
-1 & \;\;\,1 & -1 & \;\;\,1 \\  
-1 & -1 &\;\;\,1 & \;\;\,1 \\ 
\;\;\,-1 & \;\;\,-1 & \;\;\,-1 & \;\;\,-1
\end{array}
\right)   
\eeq 
the  electric ($x$)  and magnetic ($y$) charges are non vanishing only for 
$(ij)$ equal to
$(12)$,$(34)$,$(56)$ and  $(78)$. From eq.s (\ref{chargeqp}) and (\ref{zZ}) we
 may read off the 
values of the charges $x_{ij}$ and $y^{ij}$:
\beqa  
x_{78}+ i y^{78}&\equiv&\frac{1}{\sqrt{2}}(Z_1+Z_2+Z_3+Z_4)\;=\; 0+i q_0 \nn \\
x_{12}+ i y^{12}&\equiv&-\frac{1}{\sqrt{2}}(Z_1-Z_2-Z_3+Z_4)\;=\;
-\frac{q_1p_1}{p_1+p_2}-ip_1 \nn \\  
x_{34}+ i y^{34}&\equiv&-\frac{1}{\sqrt{2}}(-Z_1+Z_2-Z_3+Z_4)\;=\;
\frac{q_1p_1}{p_1+p_2}-ip_2\nn \\  
x_{56}+ iy^{56}&\equiv&-\frac{1}{\sqrt{2}}(-Z_1-Z_2+Z_3+Z_4) \;=\;0-ip_3
\label{si}
\eeqa  
The relation between this representation of the dressed charges and the one in 
eq.(\ref{ourdress}), which is related to the choice of the $STU$ model in which the 
generating solution has been worked out, is the following:
\begin{eqnarray}
\{y^{12},y^{34},y^{56},y^{78}\}&=&\{y^1,y^2,y^3,-x_0\}\nonumber\\
\{x_{12},x_{34},x_{56},x_{78}\}&=&\{x_1,x_2,x_3,y^0\}
\label{corr}
\end{eqnarray} 
where symplectic transformation  $x_0=-y^{78}$, $y^0=x_{78}$, as discussed in 
\cite{mm}, is related to the feature of our $STU$ model (both in type IIA and 
type IIB cases) of being embedded {\it non--perturbatively} in the larger 
$N=8$ theory and therefore is required in order for the truncation to be 
consistent\footnote{Indeed the dimensionally reduced R--R 
vector $A_\mu$ in the $STU$ model is an electric potential, while it is 
magnetic in the $N=8$ from the type IIA viewpoint, see table 3 of \cite{mm}.}. 
The actual D0--brane effective charge is thus $y^{78}=q_0$.

The precise correspondence between the dressed charges (in the two 
representations) and the parameters associated with the microscopic 
configurations previously discussed (that is those characterizing the type 
IIB configuration of table \ref{ND3brane}: $N_0,N_1,N_2,N_3,\theta$) is represented 
in table \ref{IIAIIB}.

\begin{table} [ht]
\begin{center}
\begin{tabular}{|c|c|c|c|c|}
\hline type IIB D--branes & Charge & type IIA D--branes & & \\
\hline 3--brane(468) & 0  & 6--brane  & $y^0$ & $x_{78}$\\
\hline 3--brane(568) & $-N_1$   & 4--brane(6789)  & $y^1$ & $y^{12}$  \\
\hline 3--brane(478) & $-N_2$   & 4--brane(4589)  & $y^2$ & $y^{34}$ \\   
\hline 3--brane(469) & $-N_3 \cos^2\theta$  & 4--brane(4567) & $y^3$& $y^{56}$   \\
\hline 3--brane(579) & $N_0+\sin^2\theta\,N_3$  & 0--brane  & $-x_0$  & $y_{78}$\\ 
\hline 3--brane(479) & $\sin\theta\cos\theta\, N_3$   & 2--brane(45) & $x_1$ & $x_{12}$ \\   
\hline 3--brane(569) & $-\sin\theta\cos\theta\, N_3$  & 2--brane(67)  & $x_2$& $x_{34}$ \\  
\hline 3--brane(578) & 0  & 2--brane(89)  & $x_3$ & $x_{56}$ \\  
\hline
\end{tabular}
\end{center}
\caption{{\small The correspondence between type IIB and type IIA charges on the 
different cycles of the compactifying torus. Notice how a $\theta\not= 0$ contribution 
induces D2--brane and D0--brane effective charges while for $\theta=0$ one gets a four 
parameter configuration.}}
\label{IIAIIB}
\end{table}

Finally, according to relations (\ref{ourdress}) or (\ref{si}) and table \ref{IIAIIB} we finally
get the precise  macroscopic/microscopic correspondence\footnote{Our normalizations 
are the following. In general the $4$ dimensional charge of
a wrapped  D$p$--brane is $Q_p\,=\,\hat \mu_p \cdot V_p/\sqrt{V_6}$
where  $\hat \mu_p\,=\,\sqrt{2\pi}(2\pi\sqrt{\alpha'})^{3-p}$ is the
normalized D$p$--brane charge  density in ten dimensions. Provided
the asymptotic values of the dilatons $b_i(r)$, which parameterize  the radii
of the compactifying torus and which has been taken to be unitary (see previous 
section), it turns out that,  in units where $\alpha'=1$, the four 
dimensional fundamental quanta of charge for {\it any} kind of (wrapped)
D$p$--brane is equal to $\sqrt{2\pi}$ and our quantized charges
$(p_\Lambda,q_\Lambda)$ have  been taken in units of that quanta,
i.e. they are integer valued.}:
\beqa
N_0 = q_0-\frac{(q_1p_1)^2}{p_3(p_1+p_2)^2}\;,\; N_1= p_1\;,\;
N_2= p_2\;,\;N_3\,\cos^2\theta= p_3\;,\;\tan\theta =-\frac{q_1p_1}{p_3(p_1+p_2)} \nn \\ 
\label{mM}
\eeqa
Through equations (\ref{mM}) all the microscopic parameters, namely 
$N_0\,,\,N_1\,,\,N_2\,,\,N_3$ and the 
angle $\theta$, are expressed in terms of the quantized charges 
$(p_{\Lambda},q_{\Lambda})$ characterizing the macroscopic solution and 
this  finally allows us to characterize ``quantitatively'' its microscopic structure.
An alternative microscopic system which could reproduce our generating solution can 
be given from the $M$--theory point of view and 
consists of three M5--branes with magnetic flux on their world volumes and
intersecting on a (compact) line along which $N_0$ units of momentum
has been put.

With the above definitions, the expression of the $E_{7(7)}$ quartic 
invariant $J_4$ in the $(y^{ij},x_{ij})$ basis,\cite{kk}, is:  
{\small 
\beqa
&&J_4=-4 \left(x_{78}x_{12}x_{34}x_{56}+y^{78}y^{12}y^{34}y^{56} \right) - 
\left(x_{78}y^{78} + x_{12}y^{12}+ x_{34}y^{34}+ x_{56}y^{56}\right)^2 \nn \\
&&+\,4 \left(x_{78}y^{78}x_{12}y^{12} + x_{78}y^{78}x_{34}y^{34} + x_{78}y^{78}x_{56}y^{56} + 
x_{12}y^{12}x_{34}y^{34} + x_{12}y^{12}x_{56}y^{56}+x_{34}y^{34}x_{56}y^{56}\right)
\nonumber\\
\label{ms}
\eeqa}
and consequently, upon use of table \ref{IIAIIB}, one can easily work out the expression of  
the entropy $S=\pi\sqrt{J_4}$ written in terms of the microscopic parameters:  
\beqa
\label{entrmi}
S_{micro}= 2\,\pi\,\sqrt{\cos^2\theta\left[N_0N_1N_2N_3 - \frac{1}{4}
\sin^2\theta\, N^2_3\left(N_1-N_2\right)^2\right]} 
\eeqa 
A derivation of the above formula via microscopic counting techniques should be
performed extending the analysis of \cite{msw,moooa} to tori (also 
the results of 
\cite{per} could possibly shed some light in this direction). However, we do not try to perform 
it here. Let us just notice that for $\theta = 0$  one recovers the usual entropy 
of the four parameters solution whose derivation via microscopic counting has been 
carried out, for instance, in \cite{bala1,bm,mal}.

\section{Discussion}

In the present paper we have worked out the generating solution 
of four dimensional $N=8$  BPS black  holes in a form which could be 
easily described, applying the results of \cite{mm}, in terms of 
pure D--brane configurations upon toroidal compactification of string (or $M$) 
theory. As a result we were able to  ``pinpoint'' the precise correspondence 
between the microscopic parameters characterizing one of these configurations 
and the supergravity parameters entering the macroscopic description of the solution.
 
The relevance of this achievement relies on the possibility on one hand to 
reconstruct the whole 56--parameter $U$--duality orbit of $N=8$ BPS black 
holes, by acting on our solution by means of $E_{7(7)}$  transformations, and 
on the other hand to study in a precise fashion the action of dualities on their 
corresponding microscopic realizations. Starting from the type
IIA configuration described in the previous section and performing a
$T$--duality transformation on the whole  $T^6$,  one ends up, for instance, with 
a configuration made of $N_0$ D6--branes, 3 bunches of ($N_1,N_2,N_3$)
D2--branes along the planes ($45$),($67$),($89$) plus effective D4--brane 
charge. But, more generally, we may also unravel the microscopic properties of 
pure NS--NS black hole solutions in the same orbit, starting from the 
corresponding embedding of the $STU$ model defined in \cite{mm}, or even of 
mixed NS--NS/R--R solutions. The important point is that having now {\it both} a 
macroscopic and a microscopic description of the generating solution one can 
follow its trasformation throughout the full $U$--duality orbit. 

In this respect a challenging problem is to recover the expression of eq.(\ref{entrmi}) 
from a microscopic entropy counting point of view, performed on the corresponding D--brane 
configuration. Knowing how to act on it by means of $U$--duality can help to shed some 
light on the actual microscopic degrees of freedom of general BPS black holes, since 
the generating solution encodes, by definition, all of them. This project is left 
for future work. 

\vskip 10pt
\noindent 
{\bf Acknowlodgements}

We would like to thank M. Bill\`o, P. di Vecchia, P. Fr\`e, T. Harmark and 
N. Obers for discussions and R. Russo and C. Scrucca for useful email 
correspondence. We are also greatful to the organizers of the TMR Torino school 
on ``String  Theory and Branes physics'' during which this work has been brought 
to an end. We acknowledge partial support by ECC under contracts ERBFMRX-CT96-0045 
and ERBFMRX-CT96-0012.

\appendix
\section*{Appendix A}
\label{appendiceB}
\setcounter{equation}{0}
\addtocounter{section}{1}

The general expression of the $N=8$ central charge eigenvalues $Z_\alpha$ in terms of 
the scalar fields and the charges characterizing the $STU$ model can be worked out making 
explicit the first order equations (\ref{eqs122}) taking into account relations (\ref{stun8}). 
Following \cite{bft} we have the following:
{\small
\begin{eqnarray}
\label{mammamia}
Re Z_1&=&
\frac{-1}{2\sqrt{- 2\,b_1b_2b_3}}\left({b_1q_1} - {b_2q_2} - {b_3q_3} +
  \left( {a_2}\,{a_3}\,{b_1}  -
     {a_1}\,{a_3}\,{b_2} - {a_1}\,{a_2}\,{b_3} -
     {b_1}\,{b_2}\,{b_3} \right) \,{p_0} + \right.\nonumber \\
&&  \left. + \left( {a_3}\,{b_2}  + {a_2}\,{b_3} \right) \,
   {p_1} + \left( {a_1}\,{b_3} - {a_3}\,{b_1} \right) \,
   {p_2} + \left( {a_1}\,{b_2} - {a_2}\,{b_1} \right) \,{p_3}\right) \nonumber\\
Im Z_1&=&\frac{1}{2\sqrt{- 2 \,b_1b_2b_3}}\left(
{a_1q_1} + {a_2q_2} + {a_3q_3} +
  \left( {a_1}\,{a_2}\,{a_3} + {a_3}\,{b_1}\,{b_2} +
     {a_2}\,{b_1}\,{b_3} - {a_1}\,{b_2}\,{b_3} \right) \,
   {p_0} +\right. \nonumber \\
&& \left.- \left( {a_2}\,{a_3}  - {b_2}\,{b_3}
      \right) \,{p_1}
 - \left( {a_1}\,{a_3} + {b_1}\,{b_3} \right)
     \,{p_2} - \left( {a_1}\,{a_2} + {b_1}\,{b_2} \right) \,
   {p_3} + {q_0}\right) \nonumber\\
Re Z_2&=& (1,2,3) \rightarrow (2,1,3) \nonumber\\
Im Z_2&=& (1,2,3) \rightarrow (2,1,3) \nonumber\\
Re Z_3&=& (1,2,3) \rightarrow (3,2,1) \nonumber\\
Im Z_3&=& (1,2,3) \rightarrow (3,2,1) \nonumber\\
Re Z_4&=&\,\frac{-1}{2\sqrt{- 2\, b_1b_2b_3}}\left( {b_1q_1} + {b_2q_2} + {b_3q_3} +
  \left( {a_2}\,{a_3}\,{b_1} + {a_1}\,{a_3}\,{b_2} +
     {a_1}\,{a_2}\,{b_3} - {b_1}\,{b_2}\,{b_3} \right) \,
   {p_0} +\right. \nn \\
&&\left. - \left( {a_3}\,{b_2} + {a_2}\,{b_3} \right) \,
   {p_1} - \left( {a_3}\,{b_1} + {a_1}\,{b_3} \right) \,
   {p_2} - \left( {a_2}\,{b_1} + {a_1}\,{b_2} \right) \,
   {p_3}\right) \nonumber \\
Im Z_4&=&\frac{1}{2\sqrt{- 2 \,b_1b_2b_3}}\left({a_1q_1} + {a_2q_2} + {a_3q_3} +
  \left( {a_1}\,{a_2}\,{a_3} - {a_3}\,{b_1}\,{b_2} -
     {a_2}\,{b_1}\,{b_3} - {a_1}\,{b_2}\,{b_3} \right) \,
   {p_0} +\right. \nonumber \\
&&\left. - \left( {a_2}\,{a_3} - {b_2}\,{b_3} \right) \,
   {p_1} - \left( {a_1}\,{a_3} - {b_1}\,{b_3} \right) \,
   {p_2} - \left( {a_1}\,{a_2} - {b_1}\,{b_2} \right) \,
   {p_3} + {q_0}\right)
\end{eqnarray}
}
where it is meant that all axions and dilatons are $r$--dependent, i.e. $a_i=a_i(r)$ 
and $b_i=b_i(r)$.



\begin{thebibliography}{99}

\bibitem{strvaf} A. Strominger and C. Vafa, Phys. Lett. {\bf B379}
(1996) 99.

\bibitem{malstr} J. Maldacena and A. Strominger, Phys. Rev. Lett. {\bf
77} (1996) 428.

\bibitem{jkm} C.V. Johnson, R.R. Khuri and R.C. Myers,
Phys. Lett. {\bf B378} (1996) 78.

\bibitem{lowe} D. Kaplan, D. Lowe, J. Maldacena and A. Strominger,
Phys. Rev. {\bf D55} (1997) 4898.

\bibitem{msw} J. Maldacena, A. Strominger and E. Witten, JHEP {\bf 12}
(1997) 2.

\bibitem{str1} A. Strominger, JHEP {\bf 02} (1998) 009.

\bibitem{porn} P. Claus, M. Derix, R. Kallosh, J. Kumar,
P. K. Townsend and  A. Van Proeyen, Phys. Rev. Lett. {\bf 81} (1998)
4553.

\bibitem{mand} J.R. David, G. Mandal and S.R. Wadia, Nucl. Phys. {\bf
B544} (1999) 590.

\bibitem{carlip} S. Carlip, Phys. Rev. Lett. {\bf 82} (1999) 2828.

\bibitem {gt} G.W. Gibbons and P.K. Townsend, Phys. Lett. {\bf B454}
(1999) 187.

\bibitem{moooa}G. Lopes Cardoso, B. de Wit and T. Mohaupt, {\it
``Macroscopic entropy  formulae and non-holomorphic corrections for
supersymmetric black holes''}, hep-th/9906094;  {\it ``Area law
corrections from state counting and supergravity''}, hep-th/9910179.

\bibitem{finn1}F. Larsen and E. Martinec, {\it ``Currents and Moduli
in the (4,0) theory''},  hep-th/9909088.

\bibitem{cve1} M. Cveti\~c and A. Tseytlin, Phys. Rev. {\bf D53}
(1996) 5619;  Erratum-ibid. {\bf D55} (1997) 3907.  
\bibitem{cve2} M. Cveti\~c and D. Youm, Nucl. Phys. {\bf B472} (1996)
249.

\bibitem{bft2} M. Bertolini, P. Fr\`e and M. Trigiante,
Class. Quant. Grav. {\bf 16} (1999) 2987. 

\bibitem{bala2} V. Balasubramanian, {\it ``How to Count the States of
Extremal Black Holes in $N=8$ Supergravity''}, hep-th/9712215.

\bibitem{mm} M. Bertolini and M. Trigiante, {\it ``Regular R--R and
NS--NS BPS black holes''}, hep-th/9910237. To appear on Int.J.Mod.Phys.

\bibitem{mp1} L. Andrianopoli, R. D'Auria, S. Ferrara, P. Fr\`e and
M. Trigiante, Nucl. Phys. {\bf B509} (1998) 463.

\bibitem{bft} M. Bertolini, P. Fr\`e and M. Trigiante,
Class. Quant. Grav. {\bf 16} (1999) 1519.

\bibitem{cvhu} M. Cveti\~c and C. M. Hull, Nucl. Phys. {\bf B480}
(1996) 296.

\bibitem{shma} M. Shmakova, Phys. Rev. {\bf D56} (1997) 540.
 

\bibitem{behr} K. Behrndt, D. Lust and W.A. Sabra, Phys. Lett. {\bf
B418} (1998) 303.

\bibitem{cve3} K. Behrndt, M. Cveti\~c and W.A. Sabra, Phys. Rev. {\bf
D58} (1998) 084018.

\bibitem{ferka} S. Ferrara and R. Kallosh, Phys. Rev. {\bf D54} (1996)
1514.

\bibitem{kal1} M.J. Duff, J.T. Liu and J. Rahmfeld, Nucl. Phys. {\bf
B459} (1996) 125;  K. Behrndt, R. Kallosh, J. Rahmfeld, M. Shmakova
and W.K. Wong, Phys. Rev. {\bf D54} (1996) 6293.

\bibitem{sabra1} K. Behrndt, G.L. Cardoso, B. de Wit, D. Lust,
T. Mohaupt  and W. A. Sabra, Phys. Lett. {\bf B429} (1998) 289.

\bibitem{fermal} S. Ferrara and J. Maldacena, Class. Quantum Grav. {\bf 15} 
(1998) 749.

\bibitem{fer} S. Ferrara, R. Kallosh and A. Strominger,
Phys. Rev. {\bf D52} (1995) 5412; A. Strominger, Phys. Lett. {\bf
B383} (1996) 39; S. Ferrara and  R. Kallosh, Phys. Rev. {\bf D54}
(1996) 1525.

\bibitem{solv} L. Andrianopoli, R. D'Auria, S. Ferrara, P. Fr\`e, R. Minasian and 
M. Trigiante, Nucl. Phys. {\bf B493} (1997) 249.

\bibitem{doug} M. Douglas, {\it ``Branes within Branes''}, hep-th/9512077; 
M. Green, J.A. Harvey and G. Moore, Class. Quant. Grav. {\bf 14} (1997) 47.

\bibitem{jab} N. Ohta and P.K. Townsend Nucl. Phys.Lett. {\bf B418} (1998) 77.

\bibitem{bala0} V. Balasubramanian and R. Leigh, Phys. Rev. {\bf D55} (1997) 6415.

\bibitem{li} M. Li, Nucl. Phys. {\bf B460} (1996) 351.

\bibitem{ck} C.G. Callan and I.R. Klebanov, Nucl. Phys. {\bf B465} (1996) 473.

\bibitem{gar} J.C. Breckenridge, G. Michaud and R.C. Myers, Phys. Rev. {\bf D55} 
(1997) 6438; M.S. Costa and G. Papadopoulos, Nucl. Phys. {\bf B510} (1998) 217.

\bibitem{bala1} V. Balasubramanian and F. Larsen, Nucl. Phys. {\bf B478} (1996) 199; 
V. Balasubramanian, F. Larsen and R. Leigh, Phys. Rev. {\bf D} 57
(1998) 3509.

\bibitem{kk} E. Cartan, O Euvres compl'etes, Paris, Editions du Centre national 
de la Recherche Scientifique, 1984; R. Kallosh and B. Kol, Phys. Rev. {\bf D53} (1996) 5344.

\bibitem{per} M. S. Costa, M. J. Perry, Nucl. Phys. {\bf B524} (1998) 333; 
Nucl. Phys. {\bf B520} (1998) 205.

\bibitem{bm} K. Behrndt and T. Mohaupt, Phys. Rev. {\bf D56} (1997) 2206.

\bibitem{mal} J. Maldacena, Phys. Lett. {\bf B403} (1997) 20.

\end{thebibliography}
\end{document}